\documentclass[amssymb,aps,floatfix,floats,preprintnumbers,twocolumn,superscriptaddress,nofootinbib]{revtex4}
\usepackage{rcs}
\usepackage{color,graphicx}
\usepackage{dcolumn}
\usepackage{bm}
\usepackage[normalem]{ulem}
\usepackage{chngcntr}
\usepackage[thinspace,thinqspace]{SIunits}
\usepackage{amsmath}
\usepackage{natbib}
\usepackage{doi}
\usepackage{hyperref}

\newcommand{\muo}{$\mu_{\textrm{0}}$}

\newcommand{\CRA}{CeRh$_{2}$As$_{2}$}

\newcommand{\YRS}{YbRh$_{2}$Si$_{2}$}
\newcommand{\YNP}{YbNi$_{4}$P$_{2}$}
\newcommand{\Tc}{$T_{\textrm{c}}$}
\newcommand{\To}{$T_{\textrm{0}}$}

\newcommand{\TN}{$T_{\textrm{N}}$}

\begin{document}
\preprint{APS/123-QED}
\title{Origin of the non-Fermi-liquid behavior in \CRA}

\author{P. Khanenko}
\email[Corresponding author:~]{pavlo.khanenko@cpfs.mpg.de}
\author{D. Hafner}
\affiliation{Max Planck Institute for Chemical Physics of Solids, D-01187 Dresden, Germany}
\affiliation{Technical University Dresden, Institute for Solid State and Materials Physics, 01062 Dresden, Germany}
\author{K. Semeniuk}
\author{J. Banda}
\author{T. L\"uhmann}
\affiliation{Max Planck Institute for Chemical Physics of Solids, D-01187 Dresden, Germany}
\author{F. B\"artl}
\affiliation{Technical University Dresden, Institute for Solid State and Materials Physics, 01062 Dresden, Germany}
\affiliation{Hochfeld-Magnetlabor Dresden (HLD-EMFL) and Würzburg-Dresden Cluster of Excellence
ct.qmat, Helmholtz-Zentrum Dresden-Rossendorf, 01328 Dresden, Germany}
\author{T. Kotte}
\affiliation{Hochfeld-Magnetlabor Dresden (HLD-EMFL) and Würzburg-Dresden Cluster of Excellence
ct.qmat, Helmholtz-Zentrum Dresden-Rossendorf, 01328 Dresden, Germany}
\author{J. Wosnitza}
\affiliation{Technical University Dresden, Institute for Solid State and Materials Physics, 01062 Dresden, Germany}
\affiliation{Hochfeld-Magnetlabor Dresden (HLD-EMFL) and Würzburg-Dresden Cluster of Excellence
ct.qmat, Helmholtz-Zentrum Dresden-Rossendorf, 01328 Dresden, Germany}
\author{G. Zwicknagl}
\affiliation{Max Planck Institute for Chemical Physics of Solids, D-01187 Dresden, Germany}
\affiliation{Institute for Mathematical Physics, Technische Universit\"at Braunschweig, D-38106 Braunschweig, Germany}
\author{C. Geibel}
\affiliation{Max Planck Institute for Chemical Physics of Solids, D-01187 Dresden, Germany}
\author{J. F. Landaeta}
\affiliation{Max Planck Institute for Chemical Physics of Solids, D-01187 Dresden, Germany}
\affiliation{Technical University Dresden, Institute for Solid State and Materials Physics, 01062 Dresden, Germany}
\author{S. Khim}
\affiliation{Max Planck Institute for Chemical Physics of Solids, D-01187 Dresden, Germany}
\author{E. Hassinger}
\affiliation{Max Planck Institute for Chemical Physics of Solids, D-01187 Dresden, Germany}
\affiliation{Technical University Dresden, Institute for Solid State and Materials Physics, 01062 Dresden, Germany}
\author{M. Brando}
\email[Corresponding author:~]{manuel.brando@cpfs.mpg.de}
\affiliation{Max Planck Institute for Chemical Physics of Solids, D-01187 Dresden, Germany}
\date{\today}
\begin{abstract}
Unconventional superconductivity in heavy-fermion systems appears often near magnetic quantum critical points (QCPs). This seems to be the case also for \CRA\ (\Tc\ $\approx$ 0.31\,K). \CRA\ shows two superconducting (SC) phases, SC1 and SC2, for a magnetic field along the $c$ axis of the tetragonal unit cell, but only the SC1 phase is observed for a field along the basal plane. Furthermore, another ordered state (phase-I) is observed below $T_0 \approx 0.48$\,K whose nature is still unclear: Thermodynamic and magnetic measurements pointed to a non magnetic multipolar state, but recent $\mu$SR and NQR/NMR experiments have clearly detected antiferromagnetic (AFM) order below \To. Also, quasi-two-dimensional AFM fluctuations were observed in NMR and neutron-scattering experiments above \To. The proximity of a QCP is indicated by non-Fermi-liquid (NFL) behavior observed above the ordered states in both specific heat $C(T)/T \propto T^{-0.6}$ and resistivity $\rho(T) \propto \sqrt{T}$. These $T$-dependencies are not compatible with any generic AFM QCP. Because of the strong magnetic-field anisotropy of both the SC phase and phase I, it is possible to study a field-induced SC QCP as well a phase-I QCP by varying the angle $\alpha$ between the field and the $c$ axis. Thus, by examining the behavior of the electronic specific-heat coefficient $C(T)/T$ across these QCPs, we can determine which phase is associated with the NFL behavior. Here, we present low-temperature specific-heat measurements taken in a magnetic field as high as 21\,T applied at several angles $\alpha$. We observe that the NFL behavior does very weakly depend on the field and on the angle $\alpha$, a result that is at odd with that observations in standard magnetic QCPs. This suggests a nonmagnetic origin of the quantum critical fluctuations.
\end{abstract} 

\maketitle

\section{Introduction}
Unconventional superconductivity is often found in the vicinity of quantum critical points (QCPs). This is well established and true not just for magnetic QCPs, but also for structural QCPs~\cite{goh2015,gruner2017}, multipolar QCPs~\cite{sumita2020} and possibly for other kinds of more exotic QCPs.
\begin{figure}[hb!]
	\begin{center}
		\includegraphics[width=\columnwidth]{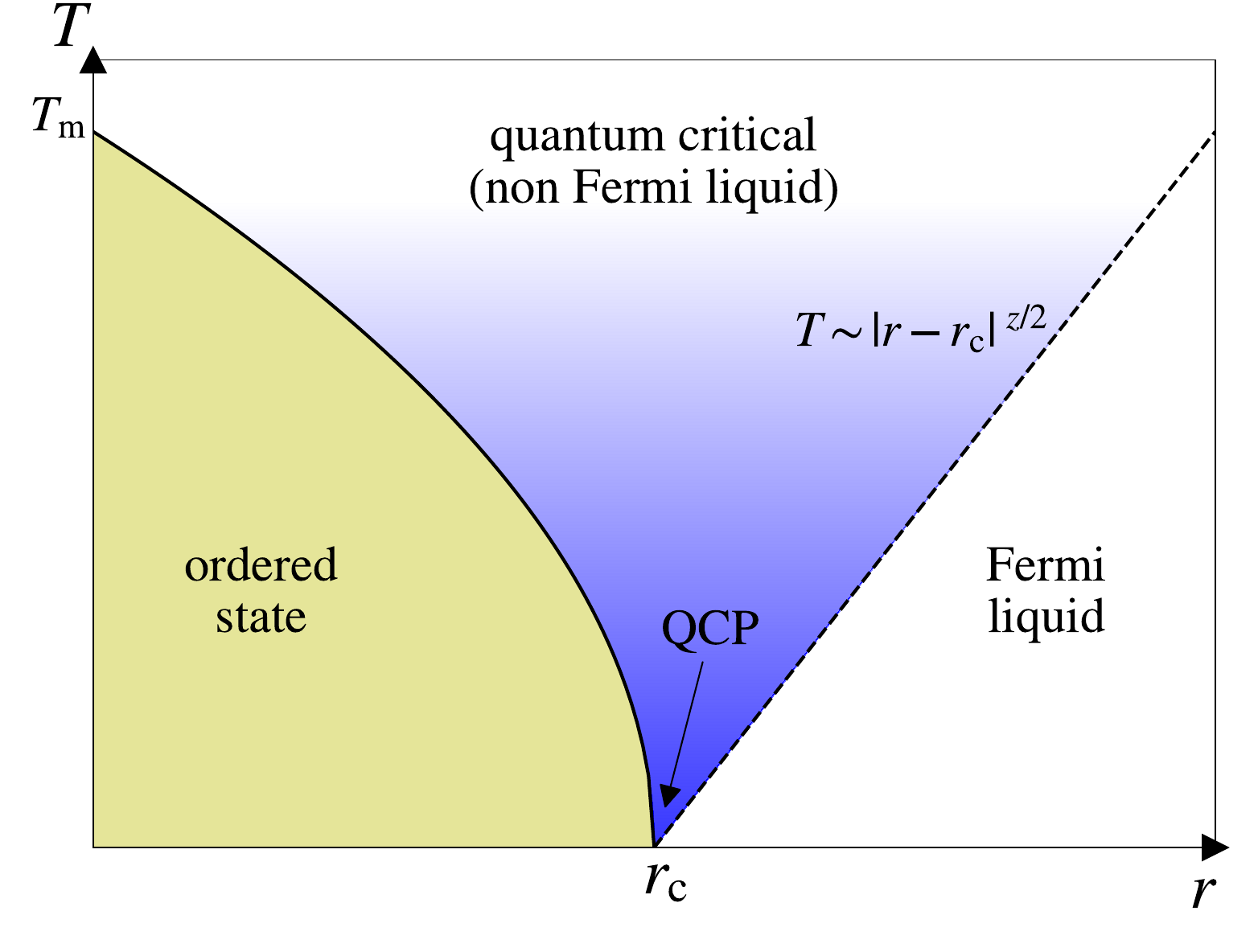}
		\caption{Schematic control parameter ($r$) - temperature $(T)$ phase diagram in the vicinity of a QCP.  The solid line indicates the finite-temperature boundary between the ordered and disordered phases. Across this line the critical behavior is typically classical. The dashed line follows $T \propto |r - r_{c}|^{\nu z}$ (see text) and indicates the crossover between the 'quantum critical region' (purple area) where non-Fermi-liquid behavior is present and the quantum disordered state in which Fermi-liquid behavior is observed. Figure adapted from Refs.~\cite{zuelicke1995,vojta2003}.}
		\label{fig1}
	\end{center}
\end{figure}
This concept is not restricted to a single class of materials, but is found in many different systems such as cuprate~\cite{taillefer2010}, organic~\cite{dressel2011}, heavy-fermion (HF)~\cite{mathur1998,yuan2003,mito2003,pham2006,landaeta2018}, pnictide~\cite{paglione2010}, and nickelate~\cite{pan2022} superconductors.

In HF superconductors there is growing evidence that quantum critical fluctuations associated with magnetic QCPs are responsible for the pairing mechanism which forms the superconducting (SC) state~\cite{monthoux2007}. This has been shown for antiferromagnetic (AFM) systems such as CeCu$_{2}$Si$_{2}$~\cite{stockert2011} and CeCoIn$_{5}$~\cite{kenzelmann2008} as well as for ferromagnetic (FM) systems such as URhGe~\cite{tokunaga2015}, UCoGe~\cite{ohta2008} and possibly UTe$_{2}$~\cite{tokunaga2019,nakamine2019}. In systems in which AFM and FM fluctuations compete, as in \YRS~\cite{ishida2002,hamann2019,schuberth2016} or CeFeAs$_{1-x}$P$_{x}$O~\cite{jesche2012}, the nature of the quantum critical fluctuations, which support superconductivity, is unclear. 

When microscopic evidence is missing, the nature of the quantum critical fluctuations can be deduced from the non-Fermi-liquid (NFL) behavior observed in transport and thermodynamic quantities within the `quantum critical region' of the phase diagram just above the QCP~\cite{loehneysen2007}. This is schematically illustrated in Fig.~\ref{fig1}: An ordered state can be continuously suppressed by a control parameter ($r$), which could be, e.g., external pressure ($p)$ or magnetic field ($H$). At $T = 0$, the quantum phase transition between the ordered and disordered phases takes place for $r = r_{c}$ at the QCP. The `quantum critical region' (purple in Fig.~\ref{fig1}) is limited by crossover lines (only one is shown in Fig.~\ref{fig1} for simplicity), which follow a specific power law, $T \propto |r - r_{c}|^{\nu z}$, where $\nu$ is the correlation-length exponent of the quantum transition and $z$ is the dynamical critical exponent. In systems with spatial dimensionality $d$, this results in an effective dimensionality $d_{eff} = d + z$~\cite{vojta2003}. Since $\nu = 1/2$ and $z = 2$ or 3 for 3-Dimensional AFM and FM systems, respectively, the crossover lines typically rise steeply from the QCP, quasi symmetrically to the boundary phase line of the ordered region (cf. dashed line in Fig.~\ref{fig1}). There are several examples of this general behavior~\cite{loehneysen2007,gegenwart2008,brando2016}.

Such behavior is also expected to occur in the recently discovered multiphase HF superconductor \CRA\ (\Tc\ $\approx$ 0.31\,K)~\cite{khim2021}. \CRA\ shows two SC phases, SC1 and SC2, for a magnetic field applied parallel to the $c$ axis of the tetragonal unit cell. These phases are separated by a weakly first-order phase transition at \muo$H^{*} = 4$\,T. In contrast, for a field applied along the basal plane only the SC1 phase is observed.
\begin{figure}[t]
	\begin{center}
		\includegraphics[width=\columnwidth]{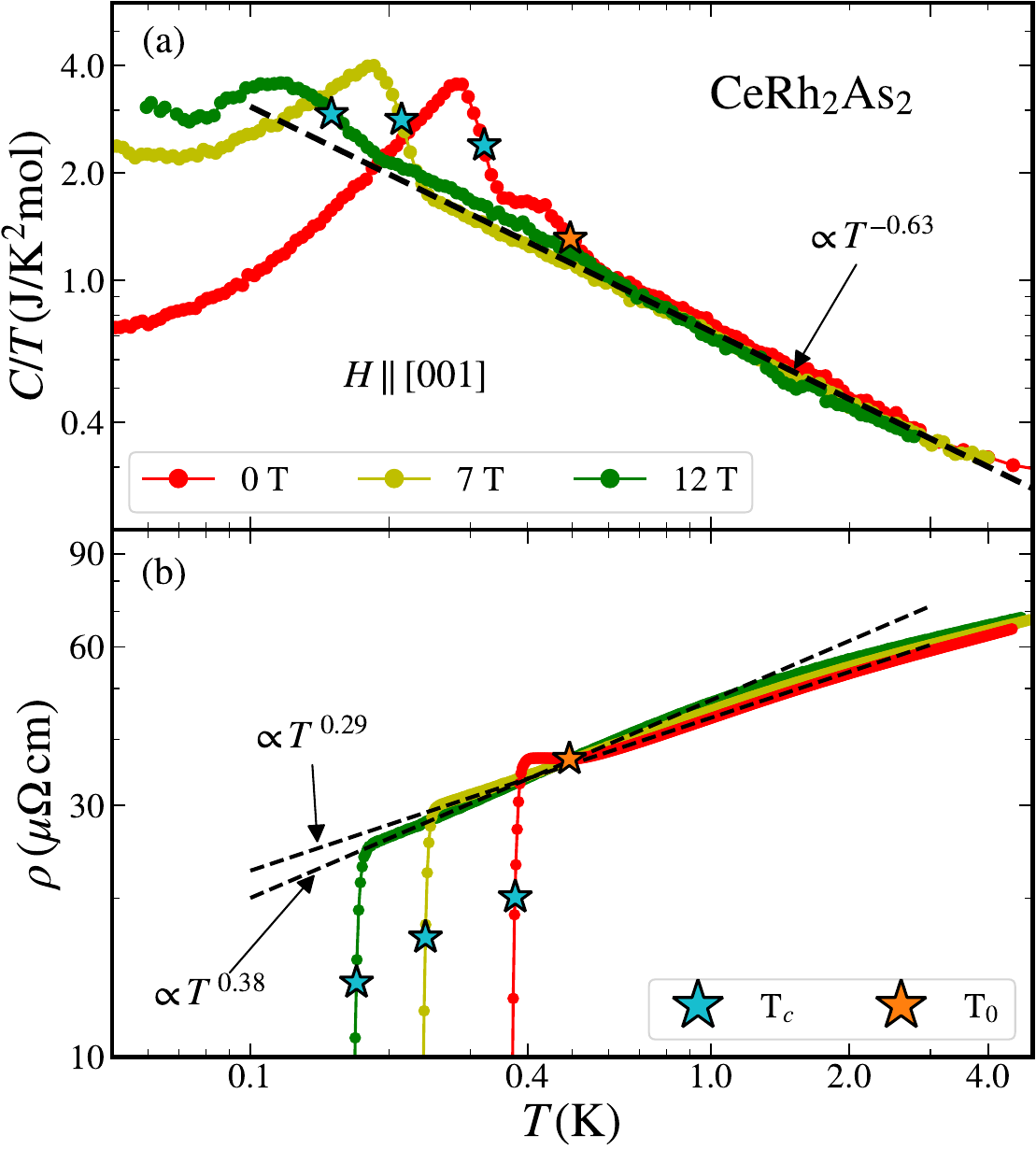}
		\caption{Electronic specific heat $C(T)/T$ and resistivity $\rho(T)$ of \CRA\ plotted as a function of temperature $T$ in double-logarithmic scale. \Tc\ and \To\ indicate the transition temperatures of the SC and phase-I orderings, respectively. The dashed lines emphasize the NFL behavior. Magnetic fields as high as 12\,T, applied along the crystallographic $c$-axis, suppress the phase-I and SC states, but the NFL behavior persists. The specific heat is plotted after subtracting the nuclear contribution due to As atoms (see Ref.~\cite{semeniuk2023}).}
		\label{fig2}
	\end{center}
\end{figure}
The anisotropic field response of superconductivity was explained by a model which takes into account a strong Rashba spin-orbit coupling due to the locally non-centrosymmetric environments of Ce sites and the quasi-2D character of the Fermi surface~\cite{yoshida2012,sigrist2014,mockli2018,schertenleib2021,skurativska2021,nogaki2021,khim2021,ptok2021,moeckli2021,nogaki2022,hafner2022,landaeta2022a,cavanagh2022,kitagawa2022,ishizuka2023,wu2023,chen2023,nally2024,lee2024}.

In addition to that, another ordered state (phase I) was observed below \To\ $\approx 0.48$\,K~\cite{khim2021}. The nature of this state is still not clear: Thermodynamic and magnetic measurements point to a non-magnetic multipolar density-wave instability supported by the presence of a Kondo-induced quasi-quartet ground state of the crystalline electric field (CEF)~\cite{hafner2022,christovam2023,takimoto2008}. This is in contrast to all known multipolar orders observed in cubic Ce-based systems, which are of local origin~\cite{effantin1985,kitagawa1996}. However, recent $\mu$SR~\cite{khim2024} and NQR/NMR~\cite{kibune2022} experiments have clearly detected AFM order below \To\ and \Tc, respectively. In the NMR experiments, it was found that antiferromagnetism is present only within the phase SC1~\cite{ogata2023,machida2022}. The relevance of dipolar degrees of freedom is supported by the analysis of the specific heat~\cite{chajewski2024}, and by NMR and inelastic neutron scattering (INS) experiments which have detected quasi-2D spin fluctuations above \To~\cite{kitagawa2022,chen2024}. The clear observation of AFM order at \To\ does not exclude the presence of quadrupolar order in \CRA: In fact, a recent theoretical analysis proposes a pure dipolar AFM state at $H = 0$ which is replaced by a mixed quadrupolar-dipolar AF state for field applied along the basal plane~\cite{schmidt2024}. An indispensible pre-requisite of the predicted quadrupolar-dipolar mixing is the admixture of a $\Gamma_{6}$ component to the local $\Gamma_{7}$ Ce 4$f$-state that is induced by the magnetic field. In \CRA, the $\Gamma_{6}-\Gamma_{7}$ mixture results from the Kondo effect.

The fact that the transition temperature \To\ is very low naturally suggets that \CRA\ is close to a QCP of phase I. In fact, recent pressure experiments reveal a suppression of phase I to a QCP at merely 0.5\,GPa~\cite{pfeiffer2024,semeniuk2024}. At ambient pressure, the presence of a QCP is based on the observation that in the normal state above the SC and I phases, \CRA\ shows a strong NFL $T$-dependence of the electronic specific heat, $C/T \propto T^{-0.63}$, and of the electrical resistivity, $\rho(T) \propto T^{n}$ with very small $n \leq 0.5$. This is shown in Fig.~\ref{fig2} for zero magnetic field and also for selected magnetic fields that suppress completely the ordered phase I and, partially, the SC state~\cite{semeniuk2023}. It is worth mentioning that these power laws can not be explained by standard theories for magnetic QCPs~\cite{loehneysen2007}. In fact, in the theory of itinerant magnetism, also known as the spin-density-wave (SDW) scenario~\cite{moriya1973,hertz1976,millis1993}, very precise expectations of thermodynamic quantities can be calculated in the Fermi-liquid and quantum critical regime~\cite{zuelicke1995,zhu2003,garst2005}. For example, $C/T \propto -\log(T)$ for a 3D ferromagnet ($z = 3$) or a 2D antiferromagnet, and $C/T \propto const-AT^{1/2}$ for a 3D antiferromagnet ($z = 2$). Only in two other HF compounds, \YRS~\cite{custers2003} and \YNP~\cite{steppke2013}, similar power laws in $C/T$ were observed at the QCP. Furthermore, the lowest resistivity exponent $n$ observed in QC systems is about $n = 1$. Theoretically, a resistivity exponent below 1 is only predicted by the two-channel Kondo model with $n = 0.5$~\cite{ludwig1991,schlottmann1993,miyake2024}. In \YRS\ and \YNP, the anomalous NFL behavior at the QCP was interpreted within the `local moment' scenario~\cite{si2001,coleman2001,si2010}, but recent investigations suggest that \YRS\ is located at a quantum tricritical point, at which both AFM and FM fluctuations diverge~\cite{misawa2008,misawa2009,hamann2019}. In \CRA, FM fluctuations can be excluded, but spatially two-dimensional AFM ﬂuctuations were detected in NMR and INS experiments~\cite{kitagawa2022,chen2024}. Finally, in \CRA\ it is worth considering the possibility of having the NFL behavior caused by a superconducting QCP as observed, e.g., in the series CeCoIn$_{5-x}$Sn$_{x}$~\cite{ramos2010}. For a SC QCP, which develops in a BCS superconductor, theory predicts $z = 2$ with similar $T$-dependencies for $C/T$ as those predicted for an AFM metal~\cite{ramazashvili1997}. Thus, all in all, the NFL behavior in \CRA\ cannot be explained by any of the models mentioned above.

In \CRA, because of the strong anisotropy of the SC and I phases with respect to the magnetic field, it is possible to tune the system to a field-induced SC QCP or a phase-I QCP by varying the angle $\alpha$ between the field and the $c$ axis.
\begin{figure*}[t]
	\begin{center}
		\includegraphics[width=\textwidth]{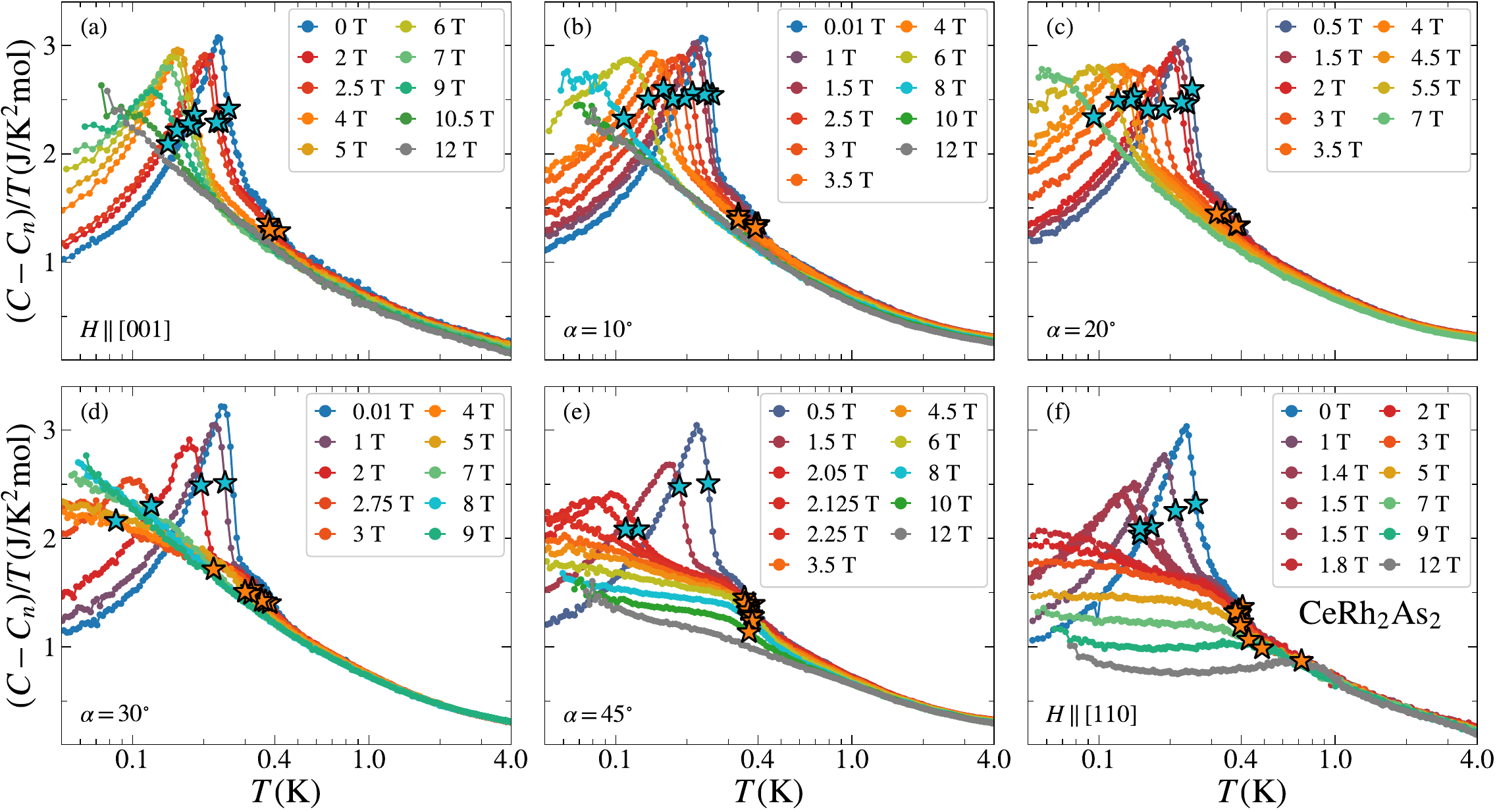}
		\caption{Temperature dependence of the specific heat with nuclear part ($C_n$) subtracted, plotted as $(C - C_{n})/T$ and measured with magnetic fields applied at six different angles $\alpha$ to the crystallographic $c$ axis. For clarity, we have plotted here only a few selected measurements, an extended set of data is provided in the Supplemental Material (SM)~\cite{sm}. The filled cyan and orange stars mark the positions of the transition temperatures \Tc\ and \To\ on the data, respectively, as in Fig.~\ref{fig2}.}
		\label{fig3}
	\end{center}
\end{figure*}
Thus, by examining the behavior of $C(T)/T$ across these QCPs, we can determine which phase is associated with the NFL behavior. Here, we present low-temperature specific-heat measurements taken in magnetic fields as high as 21\,T applied at several angles $\alpha$. Surprisingly, we observe that the NFL behavior depends very weakly on the field and on the angle $\alpha$. Although the applied magnetic fields correspond to a large energy compared to the ordering temperatures and the critical fields of the ordered phases, the NFL behavior persists with a similar power law down to the lowest temperature of about 50\,mK also at those fields. This result is at odd with that observed for standard magnetic QCPs and points to a nonmagnetic origin of the quantum critical fluctuations.
\section{Results}
The samples used in this study are listed in Table~\ref{tab}: We used samples from our first batch~\cite{khim2021,hafner2022}, but also larger and higher-quality samples (comparable to those used in Refs.~\cite{chajewski2024b,chajewski2024}) from new batches~\cite{khim2021,hafner2022,landaeta2022a}. The addenda consisted mostly of Ag wedges that we used to hold the samples on the measuring platform aligned at an angle $\alpha$ to the field: $\alpha = 0$ for $H \parallel [001]$ and 90\degree\ for $H \parallel [110]$. Since \CRA\ is a HF compound the heat capacity of the Ag wedges (similar weights as the samples) is a hundred to a thousand times smaller than that of the samples below 2\,K.
\begin{table}[b]
	\centering
	\caption{Single crystals and silver wedges used in the experiments. The angle $\alpha$ is between the field $H$ and the $c$ axis.}
	\label{tab}
	\begin{ruledtabular}
		\begin{tabular}{lllrr}
			N. & batch & mass (mg) & $\alpha$ ($\degree$) & Ag wedge (mg) \\
			\hline
			1 & 87334 (Ref.~\cite{khim2021}) & 5.06 & 0, 90 $\pm$ 1 & --\\
			2 & 87444 (Ref.~\cite{hafner2022}) & 20.04 & 90 $\pm$ 1 & --\\
			3 & 87444 & 19.00 & 10 $\pm$ 2 & 31.75\\
            3 & 87444 & 19.00 & 20 $\pm$ 2 & 59.40\\
			3 & 87444 & 19.00 & 45 $\pm$ 2 & 35.80\\
			4 & 87444 & 14.55 & 30 $\pm$ 2 & 41.30\\
            5 & 87700 (Ref.~\cite{semeniuk2023}) & 3.11 & 0 $\pm$ 1 & --\\
            6 & 87700 & 3.34 & 0 $\pm$ 1 & --\\
            7 & 87726 & 3.96 & 90 $\pm$ 1 & --\\
		\end{tabular}
	\end{ruledtabular}
\end{table}

In Fig.~\ref{fig3} we show data of the specific heat $C(T)/T$ taken with magnetic field applied at different angles $\alpha$ to the crystallographic $c$ axis (as listed in Table~\ref{tab}). These measurements were carried out in a dilution refrigerator at temperatures down to 50\,mK and to a maximum magnetic field of 12\,T. We subtracted from every measurement the nuclear contribution $C_{n}$ due to the quadrupolar and Zeeman splitting of the As nuclear spins ($I = 3/2$).
\begin{figure*}[t]
	\begin{center}
		\includegraphics[width=\textwidth]{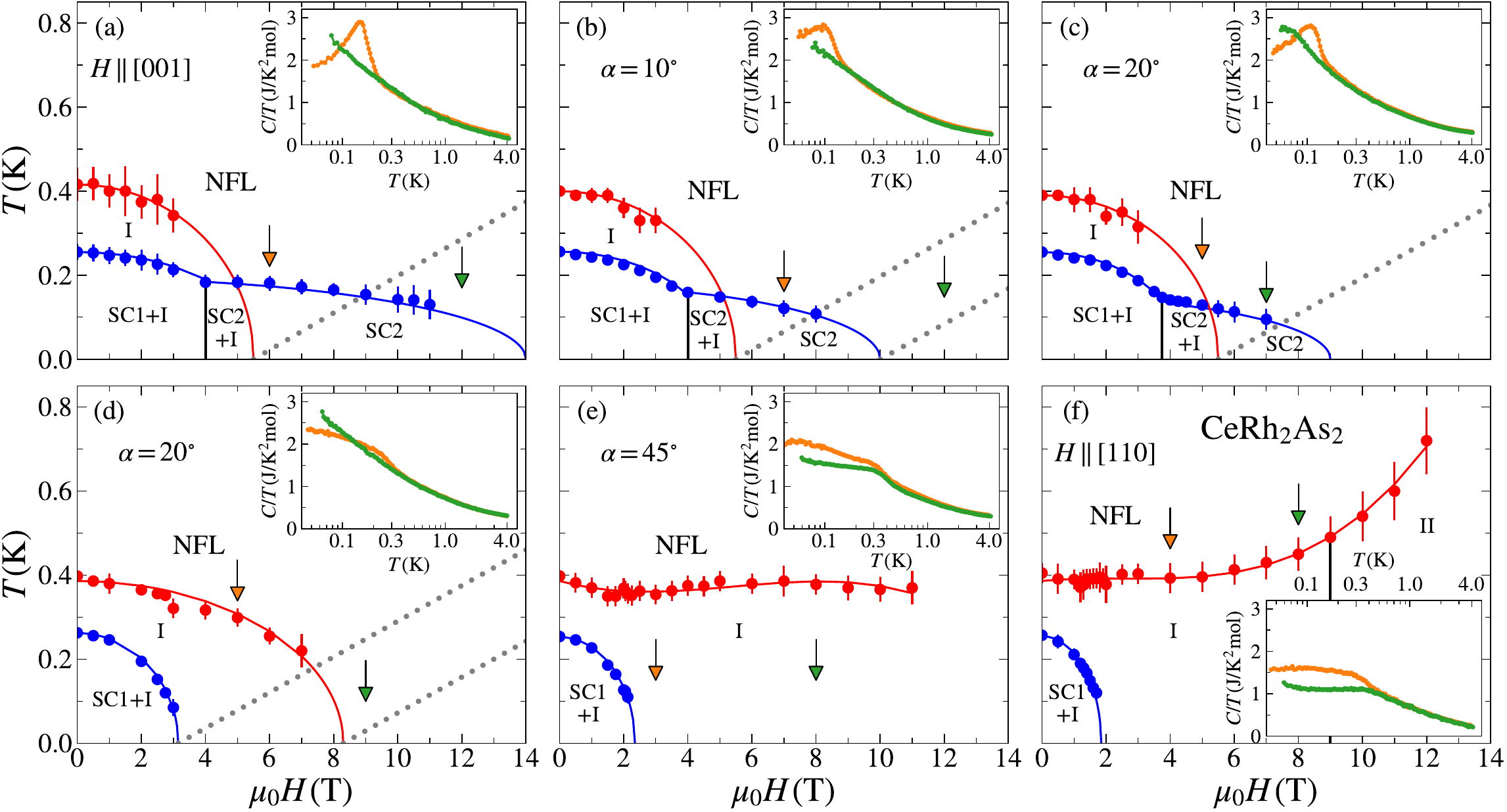}
		\caption{Temperature vs. field phase diagrams extracted from the specific-heat data shown in Fig.~\ref{fig3}. Red points indicate \To\ and blue points indicate \Tc. The solid lines are fits to the data. Only in panels (e) and (f) the solid red lines are guides for the eye. The black vertical lines in panels (a), (b), and (c) indicate the first-order phase transition from SC1 to SC2, whereas the black vertical line in panel (f) indicates the transition between phase I and phase II. The dotted lines indicate possible crossover lines from NFL behavior at high $T$ to FL behavior at low $T$ arising from different QCPs. The insets show $T$-dependent measurements at selected field marked by green and orange arrows in the main panels.}
		\label{fig4}
	\end{center}
\end{figure*}
This procedure is precisely described in the Supplemental Material of Ref.~\cite{semeniuk2023}. At zero magnetic field, we observe two second-order phase transitions at \To\ and \Tc, which indicate the transitions into the phase-I and SC states, respectively. Because both transitions have second-order character, we define the transition temperatures using the equal-entropy construction, as it was done before in Refs.~\cite{khim2021,hafner2022,semeniuk2023}. The transition points are marked by stars on the data and are plotted in the $H-T$ phase diagrams of Fig.~\ref{fig4}.

As previously shown in Ref.~\cite{landaeta2022a}, a magnetic field suppresses both \To\ and \Tc\ only for angles $\alpha \leq 30\degree$ (cf. Fig.~\ref{fig3}(a)-(d) and Fig.~\ref{fig4}). At larger angles this is not the case for \To\ anymore: At $\alpha = 45\degree$, \To\ stays almost constant for fields up to 11\,T (Figs.~\ref{fig3}(e)); for $\alpha = 90\degree$, \To\ is constant up to about 5\,T and then increases steeply with field as shown in Fig.~\ref{fig3}(f), Fig.~\ref{fig7} and Ref.~\cite{hafner2022}. This increase is accompanied by a strong first-order phase transition at about 9\,T (black vertical line in Fig.~\ref{fig4}(f)) into another high-field ordered phase, labeled `phase II' in Ref.~\cite{hafner2022}. 

What is relevant for our study is the behavior of $C(T)/T$ at high fields. In all data for $\alpha \leq 30\degree$, we notice that as soon as the ordered phases are suppressed, $C(T)/T$ continues to exhibit a very strong divergence (stronger than $-\log T$) with decreasing $T$ without showing any crossover to a constant value as expected for a FL ground state. On the other hand, for in-plane fields ($\alpha = 90\degree$) below \Tc\ and \To, i.e. within the ordered phases, $C(T)/T$ tends to recover a constant value. At $\alpha = 45\degree$ the situation is more complex: Below the transition into phase I at \To, $C(T)/T$ still increases towards lower $T$, but at a smaller rate and logarithmically. We will comment on this behavior later. From these preliminary observations we can conclude that \CRA\ shows strong NFL behavior within the normal-state region of the $H - T$ phase diagram at any angle, but shows tendency to recover a FL state within the ordered states for $T \rightarrow 0$.

A standard method to obtain insight into the origin of the NFL behavior is to compare in a phase diagram the limits of the regions with NFL behavior with the phase boundaries of the ordered states. For this reason, we examine now the $H - T$ phase diagrams deduced from the present experimental results. They are presented in Fig.~\ref{fig4}. Red points indicate \To\ and blue points indicate \Tc. For simplicity, we have extrapolated the phase boundary lines of all phases by describing the data with a function $T = T_{x}\sqrt{1 - (H/H_{x})^{2}}$, where $T_{x}$ is the transition temperature at $H = 0$ and $H_{x}$ is the critical field. Only in panels (e) and (f) the solid red lines are guides for the eye. A more precise analysis of the phase-boundary lines with a model based on Ginzburg-Landau theory of coupled order parameters~\cite{imry1975} - as done, e.g., for iron-pnictide superconductors in Ref.~\cite{fernandes2010} - was presented by us in Ref.~\cite{semeniuk2023}. While this analysis indicated a weak competing coupling between the order parameters of phase I and SC2, for simplicity in the present paper we set this interaction to zero for the fits. This has only a small effect on the predicted $T_{0}(H)$ phase boundary, which is still close to the one determined experimentally~\cite{khanenko2024}.
%
%\replace{It also implies that the phase boundary line of phase-I should enter the SC2 phase (cf. panels (a), (b) and (c) of Fig.~\ref{fig4}), changes the slope slightly and becomes so steep to remain invisible in $T$-sweep measurements of the specific heat. Recent high-resolution magnetostriction and field-dependent susceptibility measurements (on samples of better quality) clearly confirm this prediction~\cite{khanenko2024}. For simplicity, in our discussion here we leave the \To\ boundary line to enter continuously phase SC2 without changing slope, assuming negligible interaction.}{}
%

The black lines in panels (a), (b), and (c) of Fig.~\ref{fig4} are the first-order phase transition line between SC1 and SC2, which has been interpreted as the transition from even- to odd-parity superconducting order parameter~\cite{yoshida2012,sigrist2014,mockli2018,schertenleib2021,skurativska2021,nogaki2021,khim2021,ptok2021,moeckli2021,nogaki2022,hafner2022,landaeta2022a,cavanagh2022,kitagawa2022,ishizuka2023,wu2023,chen2023,nally2024,lee2024}, a topic which is not explicitly discussed here. The black line in panel (f) indicates the phase transition at a critical field of about $H_{\mathrm{cr}} \approx 9$\,T between phase I and phase II~\cite{hafner2022}. Since recent $\mu$SR experiments gave evidence of AFM ordering below \To\ and that this order coexists with superconductivity~\cite{khim2024}, we label the mixed phases below \Tc\ as SC1+I and SC2+I. This may suggest the presence of exotic intertwined order parameters~\cite{fradkin2015,aishwarya2023,gu2023,szabo2023,lee2024}. Figures~\ref{fig4}(a)-(f) evidence considerable anisotropies of the phase diagrams, with opposite angular dependencies of the SC and phase-I states. Finally, the dotted lines mimic the expected position of crossover lines (cf. Fig.~\ref{fig1}) associated to QCPs below which FL behavior is expected. For instance, in Fig.~\ref{fig4}(d) the two dotted lines are associated to the field-induced SC1+I QCP and phase-I QCP. Importantly, the slope of these lines strongly depends on the coupling between the field and the fluctuations of the order parameters, i.e., on the nature of the order parameter. For an AFM QCP we have $z = 2$ and expect a linear behavior (cf. Fig.~\ref{fig1}). For simplicity, we have drawn them to be rather symmetric with respect to the position of the QCP. This corresponds to the assumption that the external field couples with the same strength to the fluctuations of the order parameter as to the order parameter itself.
%
%\replace{This is expected, e.g., for magnetically ordered states. In \CRA, since the $4f$-electron magnetic moment is strongly reduced to values below 0.1\,\muB\ by the Kondo effect, a field of 10\,T would correspond to an energy of about 0.67\,K which is close to \To. Thus, for instance, if the field couples strongly to the order parameter of phase-I we expect to observe FL behavior in our measurements at a field close to 10\,T below a temperature of about 0.4\,K.}{}
%
\begin{figure}[ht!]
	\begin{center}
		\includegraphics[width=\columnwidth]{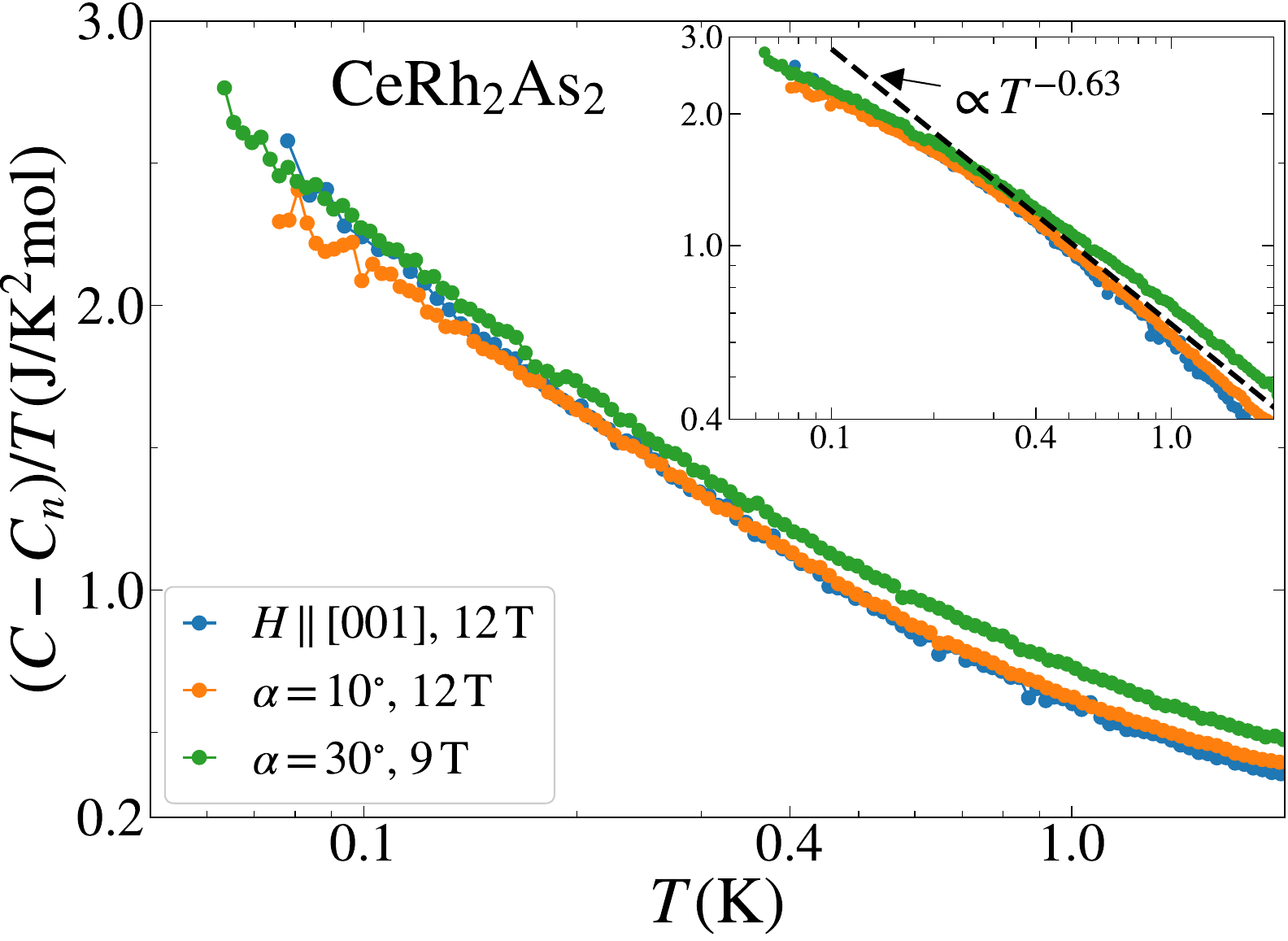}
		\caption{$(C - C_{n})/T$ measured at selected magnetic fields between 9 and 12\,T and angles between $0\degree$ and $30\degree$ within the normal state. All curves overlaps within the measured $T$-range and show larger than logarithmic divergence of $C/T$ to lower $T$. The inset shows the same data on a double-logarithmic scale to emphasize the weak deviation from the zero-field NFL behavior ($C/T \propto T^{-0.63}$) below 0.3\,K.}
		\label{fig5}
	\end{center}
\end{figure}

We take a closer look now at the $T$-dependence of $C(T)/T$ at selected fields marked by orange and green arrows in every panel of Fig.~\ref{fig4}. These data are plotted in the respective insets. The orange curves correspond to a situation with a stable SC or phase-I order, while the green curves correspond to a situation where the ordered states are either significantly or completely suppressed. In the inset of Fig.~\ref{fig4}(a) the orange data are taken at 6\,T, i.e., at the putative phase-I QCP. The specific heat shows NFL behavior in the normal state but tends to recover a constant $C/T$ for $T \rightarrow 0$ below the SC transition, which is at \Tc\ $\approx 0.2$\,K. In contrast, the green data set at 12\,T shows NFL behavior down to 0.075\,K. If we had a phase-I QCP at $H_{\mathrm{0}} = 6$\,T, then at twice this field strength one would expect a crossover to FL behavior at the lowest temperatures, typically in the range below the dotted line. However, there is no indication of such a crossover. We find a similar behavior for $\alpha = 10\degree$ shown in Fig.~\ref{fig4}(b)~\cite{footnote}. This rules out that the NFL behavior is due to fluctuations associated with the field-induced phase-I QCP. Also fluctuations associated with a field-induced SC2 QCP can be ruled out. This is because, although the SC2 phase dome shrinks to lower fields with increasing angle, at 12\,T the $T$-dependence of $C/T$ down to 0.075\,K does not change at all with angle. Even when the phase SC2 is completely gone at $\alpha = 30\degree$, $C/T$ shows the same $T$-divergence, as shown for a field of 9\,T in the inset of Fig.~\ref{fig4}(d). This also rules out a field-induced SC1+I QCP. Interestingly, the measurement close to SC1+I for $\alpha = 45\degree$ (orange curve in Fig.~\ref{fig4}(e)), shows that $C/T$ increases logarithmically below \To\ down to 0.04\,K. This increase becomes weaker with increasing field, e.g., at 8\,T (green curve in the inset of Fig.~\ref{fig4}(e)). This suggests that the order at \To\ does not suppress all QC fluctuations. A logarithmic increase of $C/T$ is expected for a 2D AFM system, which would be in agreement with 2D AFM ﬂuctuations detected in NMR and INS experiments~\cite{kitagawa2022,chen2024}. Interestingly, the effect of pressure is in one aspect very different from the effect of magnetic field. Pressure suppresses phase I, which disappears at about 0.5-0.7\,GPa~\cite{pfeiffer2024,semeniuk2024}. At the same time, it start suppresses the NFL behavior, which starts to be suppressed at about the same pressure. Thus, in contrast to magnetic field, pressure displays the expected and usually observed behavior: It suppresses both the ordered state as well as the QC fluctuations.

To summarize our main finding in a single figure, we plot three selected data sets in Fig.~\ref{fig5}. These are data already shown (green curves) in the insets of Figs.~\ref{fig4}(a),~\ref{fig4}(b), and~\ref{fig4}(d), i.e., for angles between $0\degree$ and $30\degree$. All data lie on top of each other. The inset shows that all curves follow the power law $T^{-0.63}$ down to $0.3$\,K, below which they start to deviate from this power law. This clearly indicates that the magnetic field only slightly suppresses QC fluctuations even at 12\,T, i.e., at relatively large fields for an ordering temperature of about half a kelvin. We can, therefore, conclude that the field couples only very weakly to the order parameter that generates the NFL behavior in the normal state.
\begin{figure}[t]
	\begin{center}
		\includegraphics[width=\columnwidth]{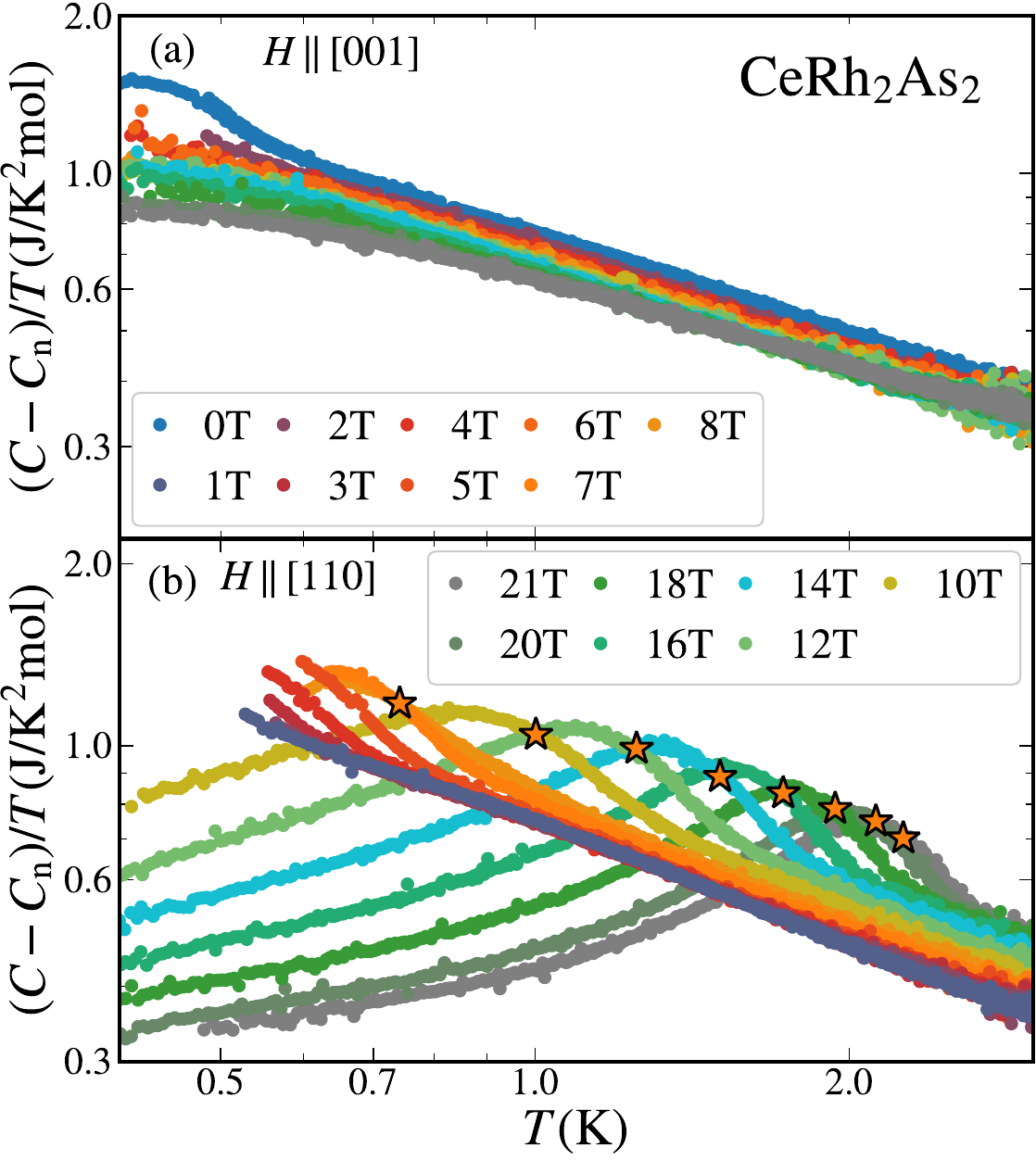}
		\caption{$(C - C_{n})/T$ of a sample from the batch 87700 measured at magnetic fields up to 21\,T. The field was applied (a) along the $c$ axis ($H \parallel [001]$) and (b) within the basal plane ($H \parallel [110]$). The filled orange stars indicate the position of the transition at \To.}
		\label{fig6}
	\end{center}
\end{figure}

The obvious question is whether in \CRA\ a magnetic field can suppress the NFL behavior at all. Therefore, we measured the specific heat using a $^{3}$He cryostat in static fields up to 21\,T. We used samples from a batch of better quality~\cite{semeniuk2023}. A summary of the results is presented in Fig.~\ref{fig6}: The field was applied within the basal plane ($H \parallel [110]$, Fig.~\ref{fig6}(b)) and along the $c$ axis ($H \parallel [001]$, Fig.~\ref{fig6}(a)). The transition temperature \To\ increases with increasing $H \parallel [110]$, as previously observed~\cite{hafner2022}, and the transition remains sharp even at the highest measured field of 21\,T. This implies that the field supports this particular order parameter as expected, e.g., for a quadrupolar~\cite{schmidt2024} or an octupolar state~\cite{matsumura2012}. Below \To\ a FL behavior is recovered in $C(T)/T$ (as already shown in the inset of Fig.~\ref{fig4}(f)) as well as in $\rho(T)$ (cf. Ref.~\cite{hafner2022}). However, for $H \parallel [001]$, the NFL behavior persists even at fields as high as 21\,T, although the divergence is weaker at this field and $C/T$ seems eventually to approach a constant value towards lowest temperatures. This further confirms the presence of a very weak coupling between magnetic field and the QC fluctuations.
\section{Discussion}
Now, we discuss possible origins of the NFL behavior in \CRA. We have shown here that the NFL behavior observed in $C(T)/T$ and in resistivity is unique. The $T$-dependent power laws $C(T)/T \propto T^{-0.63}$ and $\rho(T) \propto T^{0.38}$ (at zero field) are very different from those predicted for standard magnetic and SC QCPs~\cite{zhu2003,ramazashvili1997}. Moreover, the NFL behavior persists to very high fields and very low temperatures, so that the field seems to couple only weakly to the QC fluctuations. Only at very high fields $H \parallel [001]$, $C(T)/T$ seems to approach a constant value (FL behavior) at lowest temperature (Fig.~\ref{fig6}).

A large field scale for the suppression of NFL behavior is on its own not a surprise, since the critical field needed to suppress magnetically ordered states in HF systems can also be very high. For example, critical fields as high as 80\,T and 50\,T are necessary to suppress the AFM order in the archetypical HF systems CeIn$_{3}$~\cite{tokunaga2019} and CeRhIn$_{5}$~\cite{takeuchi2001,helm2020}, respectively. This can be simply explained: The magnetic moments are screened by the Kondo effect, and applying a magnetic field in an AFM system results in two competing effects: On the one hand, the field suppresses the AFM order as in any standard AFM system, on the other hand it stabilizes magnetic states because it suppresses the Kondo screening. Therefore, in a number of Kondo systems the effect of a magnetic field on \TN\ is essentially zero up to quite large fields. But there is, yet, no known example where this competition results in a sizeable increase of \TN, as it is observed in \CRA~\cite{hafner2022}.

What matters is the rate between the field scale $H_{NFL}$ needed to suppress the NFL behavior and that required to suppress the ordered state. In \CRA, the field scale $H_{NFL}$ is at least three times larger than the field which suppresses phase I and about two times larger than the upper critical field of SC2. Thus, there is a large field range in which the AFM order is fully suppressed, i.e., where the system is in a field-polarized FM-like state, but where the fluctuations that induce the NFL behavior remain unaffected. This cannot be explained by the competing effects mentioned above, since in the field-polarized state these two effects are cooperative instead of being competitive: Increasing the field stabilizes the field-polarized state as in any AFM system, but even more in a Kondo system through weakening of the Kondo effect. The large field range, in which NFL is observed in \CRA\ after suppression of the ordered state, is in strong contrast to the behavior observed in archetypical QC systems such as, e.g., \YRS~\cite{custers2003} and CeIn$_{3}$~\cite{knebel2001}, where the FL is recovered very close to the field- and pressure-induced QCP, respectively.

\begin{figure}[t]
	\begin{center}
		\includegraphics[width=\columnwidth]{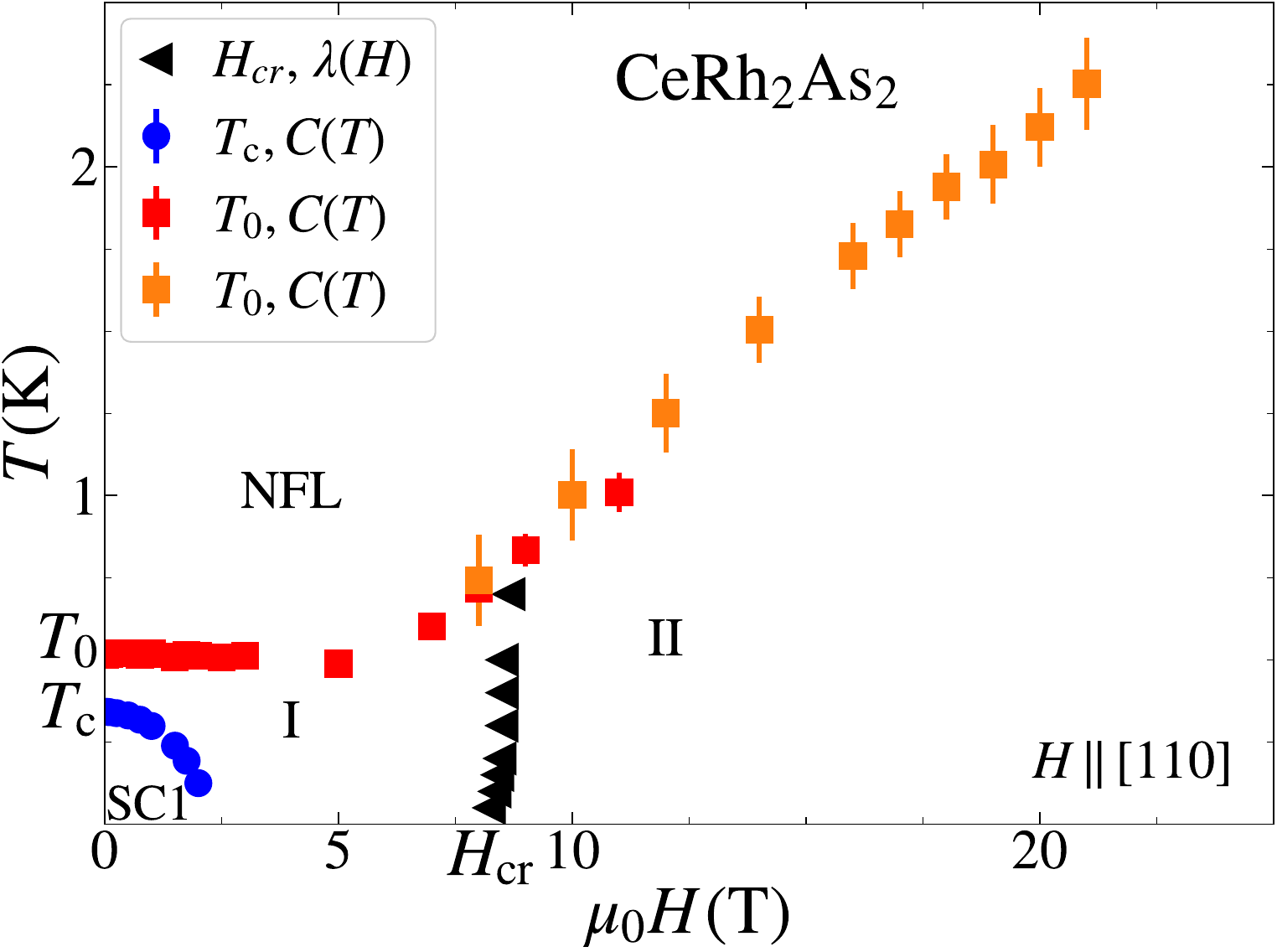}
		\caption{Extended phase diagram of \CRA\ for $H \parallel [110]$ extracted from thermodynamic measurements on samples from recent batches 87700 and 87726. $\lambda(H)$ is the magnetostriction coefficient. The red and orange squares denote data taken in a 12\,T and 21\,T system, respectively.}
		\label{fig7}
	\end{center}
\end{figure}

These observations point to the presence of non-dipolar, but multipolar QC fluctuations. The presence of a quadrupolar ordered state below \To\ in \CRA\ has been already proposed by Hafner \textit{et al.}~\cite{hafner2022}. The main experimental observation that supports such a proposal is the very anisotropic phase diagram for \To: In particular for $H \parallel [110]$ (Fig.~\ref{fig7}), \To\ increases steeply with increasing field such that \To\ $\approx 2$\,K at 20\,T, a temperature four times larger than the one at zero field. This reminds us of the case of CeB$_{6}$ for $H \parallel [110]$~\cite{goodrich2004,jang2017}, in which the antiferro-quadrupolar (AFQ) ordered state is initially stabilized by the field, with the transition temperature increasing up to 10\,K at about 30\,T. If \To\ would be due to AFM or a SDW transition, we would expect \To\ to be reduced by the field. An increasing transition temperature with magnetic field is naturally observed in ferromagnets, but FM ordering in \CRA\ is excluded because of the very small paramagnetic susceptibility and the fact that \To\ remains sharp with increasing field, whereas in ferromagnets it would broaden. There are, however, particular cases, such as 2D frustrated magnets, in which $T_{\mathrm{N}}$ initially increases with increasing field, reaches a maximum at $T_{\mathrm{max}}$ and then decreases. But the difference between $T_{\mathrm{max}}$ and  $T_{\mathrm{N}}$ is only a few percent~\cite{schmidt2017} in these systems. Another example is the non-centrosymetric superconductor CePt$_{3}$Si which shows an almost constant \TN\ in field~\cite{takeuchi2004,takeuchi2013}, similar to what we observe in \CRA\ up to 5\,T.

The evidence of AFM ordering observed in $\mu$SR and NMR experiments is, in general, not an argument against the presence of quadrupolar order or fluctuations. A recent theoretical study, which takes into account the precise CEF scheme of \CRA\ in magnetic fields, neglecting the Kondo effect, shows that the transition at \To\ could be purely dipolar (AFM), but an in-plane field would induce a quadrupolar, possibly $O_{xy}$~\cite{schmidt2024} order parameter coupled to the dipolar AFM order parameter. The authors could reproduce essential features of the observed phase diagrams of \CRA, in particular the large anisotropy and the strong increase of \To\ for larger in-plane fields. In their model the quadrupolar phase II is induced by the mixing of the first CEF excited state into the ground state. This mixing is zero at zero field, therefore the order in phase I is purely dipolar. However, as demonstrated in Ref.~\cite{hafner2022}, the Kondo effect in \CRA\ can induce a similar mixing even at zero field because the quasi-quartet energy gap is of the same order as the Kondo temperature~\cite{christovam2023}. So, we cannot exclude that also quadrupolar degrees of freedom are involved in the transition at \To\ in zero field to form a quadrupole-density-wave (QDW) ordering together with AFM order.

%\replace{The behavior of \To\ in \CRA\ reminds us also of that of CeB$_{6}$ for $H \parallel [110]$, in which an antiferro-quadrupolar (AFQ) ordered state is transformed into an octupolar ordered state in field~\cite{matsumura2012}.  The $(O_{yz}-O_{zx})$ quadrupole order parameter increases concavely from a finite value at zero field whereas the $(T_{xyz})$ octupole increases convexly from zero.  This would simply explain the phase diagram for $H \parallel [110]$ in Fig.~\ref{fig7}, as well as the phase transition observed at $H_{cr} \approx 9$\,T between phase-I and II, and finally the sharpness of the transition at all fields. In fact, the very same behavior of the transition temperature was taken as clear evidence for the existence of the AFQ interaction in CeB$_{6}$. In \CRA\ we might have a similar evolution of the multipolar order parameters, and their quantum fluctuations could then be at the origin of the NFL behavior above \To.}{}

Although rare in nature, the possibility of a mixed dipolar/quadrupolar order exists also at zero field. A clear example for that is the compound TbB$_{2}$C$_{2}$. This compound shows a strong increase of \TN\ with magnetic field, which reaches 13.5\,K for $H \parallel [110]$ at 10\,T. In early magnetization experiments it was suggested that its ground state is purely AFM and the magnetic field induces an AFQ order, but without quadrupolar order at $H = 0$~\cite{kaneko2003}. Later on, by means of resonant soft-x-ray experiments, it was directly shown that the orbitals are already ordered in the AFM phase in zero field in a dominantly ferro-quadrupolar arrangement, where the orientation of the quadrupoles is strongly tied to the direction of the AFM moments. In an external field the rotation of the AFM moments induces an increasing AFQ component in the order parameter~\cite{mulders2007}. Thus, in \CRA\ there might exist a mixed dipolar/quadrupolar order at $H = 0$ caused by the Kondo effect which mixes CEF levels even at zero field~\cite{hafner2022}. Quadrupolar ordering at such low temperature is extremely demanding to prove. Techniques such as resonant inelastic x-ray scattering, for instance, would give insight but are challenging at very low temperatures. So far, there is no direct evidence of this QDW state. Also, other more complex scenarios, such as the presence of multipolar odd-parity fluctuations are difficult to be verified~\cite{nogaki2023}.

\section{Conclusions}
From the temperature dependence of the specific heat measured at high magnetic fields and at different angles to the crystallographic $c$ axis, we deduce that the NFL behavior observed in \CRA\ cannot be explained by the presence of a standard magnetic QCP in this material, but seems to be associated with multipolar quantum fluctuations that are sensitive to pressure-tuning rather than magnetic-field tuning.
\begin{acknowledgments}
We are indebted to A. Rost, O. Stockert, B. Schmidt and P. Thalmeier for useful discussions. We acknowledge support from funding by the DFG through CRC1143 (project number 247310070) and the W\"urzburg-Dresden Cluster of Excellence on Complexity and Topology in Quantum Matter — ct.qmat (EXC 2147, project ID 390858490). S. K. acknowledges funding by the DFG through KH 387/1-1. G. Z. acknowledges that this research was supported in part by grant NSF PHY-2309135 to the Kavli Institute for Theoretical Physics (KITP). This work was supported by HLD-HZDR, member of the European Magnetic Field Laboratory (EMFL).
% This work is supported by the joint Agence National de la Recherche and DFG program Fermi-NESt. through Grants No. GE602/4-1 (C. G. and E. H.).
\end{acknowledgments}
\appendix
\bibliography{khanenko_prb_2024.bib}

\begin{thebibliography}{100}
\bibitem [{foo()}]{footnote}%
  \BibitemOpen
  \href@noop {} {\bibinfo {title} {Because the phase diagram for $\alpha =
  20\degree$ is essentially very similar to that for $\alpha = 10\degree$, we
  did not take many measurements at high fields for $\alpha =
  20\degree$.}}\BibitemShut {Stop}%
\end{thebibliography}
\bibliographystyle{apsrev4-2}
\end{document}